\documentclass[12pt]{article}
\usepackage{setspace}
\usepackage{mathtools}
\mathtoolsset{showonlyrefs,showmanualtags}
\usepackage{fullpage}
\usepackage{outline}
\usepackage{amsmath,amssymb,amsthm}
\usepackage[pdftex]{graphicx}

\newtheorem{claim}{Claim}
\begin{document}
\doublespacing
\section{Abstract}

Given points in Euclidean space of arbitrary dimension, we prove that there exists a spanning tree having no vertices of degree greater than 3 with weight at most 1.559 times the weight of the minimum spanning tree.  We also prove that there is a set of points such that no spanning tree of maximal degree 3 exists that has this ratio be less than 1.447.  Our central result is based on the proof of the following claim:

Given $n$ points in Euclidean space with one special point $v$, there exists a Hamiltonian path with an endpoint at $v$ that is at most 1.559 times longer than the sum of the distances of the points to $v$.

These proofs also lead to a way to find the tree in linear time given the minimal spanning tree.

\section{Introduction}
\label{sec:problem}
The minimum spanning tree (MST) problem in graphs is perhaps one of the most basic problems in graph algorithms.  An MST is a spanning tree with minimal sum of edge weights.  Efficient algorithms for finding an MST are well known.

One variant on the MST problem is the bounded degree MST problem, which consists of finding a spanning tree satisfying given upper bounds on the degree of each vertex and with minimal sum of edges weights subject to these degree bounds.

In general, this problem is NP-hard \cite{NPhard}, so no efficient algorithm exists.  However, there are certain achievable results.  For undirected graphs, Singh and Lau \cite{SinghLau} found a polynomial time algorithm to generate a spanning tree with total weight no more than that of the bounded degree MST and with each vertex having degree at most one greater than that vertex's bound.  If the graph is undirected and satisfies the triangle inequality, Fekete and others \cite{network} bound the ratio of the total weight of the bounded-degree MST to that of any given tree, with a polynomial-time algorithm for generating a spanning tree satisfying the degree constraints and this ratio bound.

The Euclidean case, with vertices being points in Euclidean space and edge weights being Euclidean distances, also has a rich history.  We denote (following Chan in \cite{1.633}) by $\tau_k^d$ the supremum, over all sets of points in $d$-dimensional Euclidean space, of the ratio of the weight of the bounded degree MST with all degrees at most $k$ to the weight of the MST with no restrictions on degrees ($\tau_k^\infty$ is the supremum of $\tau_k^d$ over all $d$).  For $k=2$, the bounded-degree MST problem becomes the Traveling Salesman Problem and $\tau_2^d=2$ \cite{network}, thus making $k=3$ the first unsolved case.

Papadimitriou and Vazirani \cite{NPhard} showed that finding the degree-3 MST is NP-hard.  Khuller, Raghavachari, and Young \cite{1.67} showed that $1.104\approx (\sqrt{2}+3)/4 \leq \tau_3^2 \leq 1.5$ and $1.035< \tau_4^2 \leq 1.25$. Chan \cite{1.633} improved the upper bounds to 1.402 and 1.143, respectively.
Jothi and Raghavachari \cite{degree4} showed that $\tau_4^2 \leq (2+\sqrt{2})/3\approx 1.1381$.  $\tau_5^2=1$ since there is always an MST with maximal degree 5 or less \cite{degree5}.

These same papers also studied the problem in higher dimensions.  Khuller, Raghavachari, and Young \cite{1.67} gave an upper bound on $\tau_3^\infty$ of $5/3\approx 1.667$, which Chan \cite{1.633} improved to $2\sqrt{6}/3\approx 1.633$.  These two followed the same approach, proving these bounds on a certain ratio, which we will call $r$. $r$ is the maximum ratio between the shortest path through a collection of points starting at a special point and the size of a star centered at that point. It is conjectured to actually be 1.5.

Khuller, Raghavachari, and Young \cite{1.67} showed that $\tau_3^\infty \leq r$.  This is achieved in linear time as follows:
\begin{enumerate}
\item root the original tree
\item treating the root as $v$, find a Hamiltonian path with ratio at most $r$ through its children.
\item repeat recursively on each child.
\end{enumerate}
Each vertex then has at most 3 neighbors: two as a child and one as a parent.

We improve previous upper bounds on $r$, and thus $\tau_3^\infty$, to 1.559. The proof leads to a linear time algorithm for generating the path and thus the bounded degree tree.  Our approach is based on Chan's, but we weigh paths differently and select the number of points to remove when performing the induction based on the distances of points to $v$.

We also find, by construction, a non-trivial lower bound of about 1.447 on $\tau_3^\infty$.

In Section \ref{sec:lemmas}, we go over $r$ a bit more carefully as well as refering to a useful paper and discuss how we will use it.  In Section \ref{sec:upperbound}, we improve the upper bound on $r$ to 1.559, and in Section \ref{sec:lowerbound} we improve the lower bound on $\tau_3^\infty$ to 1.447.

\section{Preliminaries}
\label{sec:lemmas}
$r$ is properly defined as follows: 

Given point $v$ and $m$ points $a_1,a_2,\ldots,a_m$ in a Euclidean space of arbitrary finite dimension, let $\displaystyle S=\sum_{i=1}^n d(v,a_i)$ and let $L$ be the length of the shortest possible path that starts at $v$ and goes around the other points in some order (it does not go back to $v$).  Then $r$ is the supremum of the possible values of $L/S$ over all arrangements of points in any number of dimensions.  $r=1.5$ is achieved for $m=2$ in one dimension by the points $v=0,a_1=1,a_2=-1$.

We use the results of Young \cite{distsum} multiple times in order to bound certain sums of distances.
This paper deals with the maximum of weighted sums (with weights $w_{i,j}$) of lengths between $n$ points in $n-1$ dimensional Euclidean space, given that each point $a_i$ is specified as being no further than some distance $l_i$ from the origin.
\begin{equation}\label{eq: Young}
\max \left(\sum_{1\leq i<j\leq n} w_{i,j} d(a_i,a_j) \right)=\min \left(\sqrt{\sum_{1\leq i<j\leq n} \frac{w^2_{\displaystyle i,j}}{x_ix_j}}\sqrt{\sum_{i=1}^n l^2_ix_i}\sqrt{\sum_{i=1}^n x_i}\right)
\end{equation}
where the maximum is taken over all arrangements of points and the minimum is taken over all nonnegative $x_i$.

Furthermore, Young specifies a relationship between the optimal arrangement and the values of $x_i$ where equality is achieved.  Thus one can iteratively approximate the optimal arrangement using the same method as in \cite{experimental}, and then calculate $x_i$ values from it.

Whenever \eqref{eq: Young} is used to give an upper bound on some weighted sum of distances, the values for $x_i$ used are given in Appendix \ref{sec: xis}.

\section{Main proof of upper bound on $r$}
\label{sec:upperbound}
Let $r^*=1.559$. We will prove that $L\leq r^*S$ (as $L$ and $S$ are defined in the introduction), thus showing that $r<1.559$.

We will prove this by strong induction on the number of points. Given $m$ vectors $a_1, a_2,\ldots,a_m$ with norms $d_1 \geq d_2 \geq d_3 \geq \ldots \geq d_m > 0$, respectively, we will try to induct by removing $a_1,\ldots,a_n$ for various values of $n$. We will try to traverse the other points, ending at $a_{n+1}$ or $a_{n+2}$. We will then add in the removed points, projected onto a sphere, and look at the average length of a path traversing them and ending at $a_1$ or $a_2$. We will then move them out in stages, seeing how this average path length changes at each stage, in order to bound the final average path length in terms of the values $d_k$. Since the average is an upper bound on the minimum, this gives us a linear inequality on the $d_k$ which is a sufficient condition for the inductive step to work. We then use linear programming to show that one of these inequalities is satisfied and thus that induction is possible. For the algorithm, we will then follow the induction to split the points up into blocks, choose the starting and ending vertex for one block at a time, using brute force to find the shortest path that goes through all the block's points.

We start by defining $a_k=0$ and $d_k=0$ for all $k>m$. Introducing these new points does not affect the distance sum or the traversing path length, as the traversing path can go to them first.

We will prove the following claim:
\begin{claim}\label{claim}
There exist two paths $P_1$ and $P_2$ ending at $a_1$ and $a_2$, respectively, such that the average of the lengths of these paths is at most $r^*S$
\end{claim}

This clearly implies that $L\leq r^*S$.

We will proceed by strong induction on $m$. To induct, remove $a_1$ through $a_n$ (where $n\geq 3$ may vary), use the inductive hypothesis to find two paths $P_1$ and $P_2$ through the other $m-n$ points, ending at $a_{n+1}$ and $a_{n+2}$, respectively. We will then try to find four paths $Q_{11}, Q_{12}, Q_{21}, Q_{22}$ with path $Q_{ij}$ going from $a_{n+i}$ to $a_j$ and going through all points $a_1, \ldots, a_n$, so that the average length of these four paths is at most $r^*(\sum_{i=1}^n d_i)$.

We will assume that this is impossible, generate a set of conditions on the values $d_n$, then prove that one of the conditions must be violated.

\subsection{Given $n \geq 3$}
\label{subsec:given}
In this section, we will assume $n>3$ to be a given value.  We will select it in Section \ref{sec: recombine}.

\begin{figure}[htbp]
\centering
\includegraphics[scale=.5]{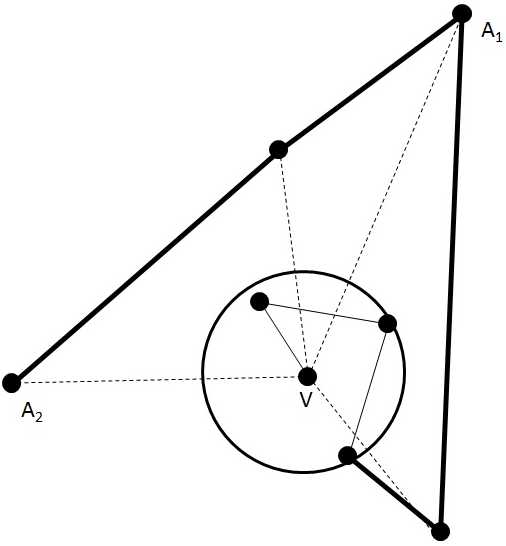}
\label{fig:thickseg}
\caption{The thick segments contribute to $L(a_n, \ldots, a_1)$; the dotted segments contribute to $S_n$.}
\end{figure}

Let $\displaystyle S_n=\sum_{i=1}^n d_i$.
Let $L(u_{n+1}, \ldots, u_1)$ be the shortest length of a path $u_{s_{n+1}}, \ldots, u_{s_1}$ where $s$ is a permutation of $1, \ldots, n+1$ so that $s_{n+1}=n+1$ and $\{s_1,s_2\}=\{1,2\}$.
Let $\overline{L(u_{n+1}, \ldots, u_1)}$ be the average length over all such paths $u_{s_{n+1}}, \ldots, u_{s_1}$. Then
\begin{equation} \label{eq: L()}
	\begin{split}
		L(u_{n+1}, \ldots, u_1)&=\frac{1}{n-1}d(u_1,u_2)+\frac{1}{2(n-1)}d(u_1,u_{n+1})+\frac{1}{2(n-1)}d(u_2,u_{n+1})\\
		+&\sum_{i=3}^n \frac{3}{2(n-1)}d(u_1,u_i)+\sum_{i=3}^n \frac{3}{2(n-1)}d(u_2,u_i)\\
		+&\sum_{i=3}^n \frac{1}{n-1}d(u_i,d_{n+1})+\sum_{3\leq i<j\leq n} \frac{2}{n-1}d(u_i,u_j)
	\end{split}
\end{equation}

We wish to find upper bounds on $\overline{L(a_{n+1}, \ldots, a_1)}$ and $\overline{L(a_{n+2},a_n,a_{n-1}, \ldots, a_1)}$. For $1\leq~i\leq~n$, let
\begin{equation} \label{eq: defineDnik}
	\begin{split}
		D_{n,i,1}&=\overline{L \left( a_{n+1}, \ldots, a_{i+1},a_i, \frac{d_i}{d_{i-1}}a_{i-1}, \ldots, \frac{d_i}{d_1}a_1 \right)} \\
		&- \overline{L \left( a_{n+1}, \ldots, a_{i+1}, \frac{d_{i+1}}{d_{i}}a_i, \frac{d_{i+1}}{d_{i-1}}a_{i-1}, \ldots, \frac{d_{i+1}}{d_1}a_1 \right)}.
	\end{split}
\end{equation}
and let
\begin{equation} \label{eq: defineDnk}
D_{n,1}=\overline{L \left( a_{n+1},\frac{d_{n+1}}{d_n}a_n, \ldots, \frac{d_{n+1}}{d_1}a_1 \right)}.
\end{equation}
$D_{n,i,2}$  and $D_{n,2}$ are defined identically, except $a_{n+1}$ and $d_{n+1}$ are replaced with $a_{n+2}$ and $d_{n+2}$.
For $k=1$ or $k=2$, 
\begin{equation} \label{eq: decompose}
\overline{L(a_{n+k},a_n,a_{n-1},\ldots,a_1)}=D_{n,k}+\sum_{i=1}^n  D_{n,i,k}
\end{equation}
Intuitively, we are setting all points at distance $d_{n+k}$, then moving out $n$ points to distance $d_n$, then moving out $n-1$ points, and so on.

We will now find values $B_{n,i}$ and $B_n$ independent of the arrangement of $a_1,a_2,\ldots$ satisfying 
\begin{equation}\label{eq: UgeqD}
	\begin{split}
		B_{n,i}(d_i-d_{i+1}) &\geq D_{n,i,1}\\
		B_n d_n &\geq D_{n,1}
	\end{split}
\end{equation}
The corresponding equations (substituting $a_{n+2}$ for $a_{n+1}$ and $d_{n+2}$ for $d_{n+1}$) will then hold for $D_{n,i,2}$ and $D_{n,2}$.

\subsubsection{$B_n$}
Define $g(n)$ as the maximum value of
\begin{align*}
	d(u_1,u_2)+\frac{1}{2}d(u_1,u_{n+1})+\frac{1}{2}d(u_2,u_{n+1})+\sum_{j=3}^n\frac{3}{2}d(u_1,u_j)+\\
	+\sum_{j=3}^n\frac{3}{2}d(u_2,u_j)+\sum_{j=3}^n d(u_j,u_{n+1})+\sum_{3\leq j<k\leq n} 2d(u_j,u_k)
\end{align*}
over unit vectors $u_1,\ldots,u_{n+1}$.

We use equation \eqref{eq: Young} to obtain upper bounds on $g(n)$, which we then use to find numerical values for $B_n$.

Substituting in $\eqref{eq: defineDnk}$ and $\eqref{eq: L()}$, we get that 
\[D_{n,1}\leq d_n\frac{g(n)}{n-1}\]
and similarly for $D_{n,2}$. Thus we can set 
\begin{equation} \label{eq:Bn}
B_n=\frac{g(n)}{n-1}
\end{equation}

\subsubsection{$B_{n,i}$}
For $i<j<k$, 
\begin{equation} \label{eq: bothstay}
	d(a_j,a_k)-d(a_j,a_k)=0.
\end{equation}
For $j\leq i< k$
\begin{equation} \label{eq: onemoves}
	d \left( \frac{d_i}{d_j}a_j,a_k \right)-d \left(\frac{d_{i+1}}{d_j}a_j,a_k \right)\leq d_i-d_{i+1}.
\end{equation}
For $k<j\leq i$
\begin{equation} \label{eq: bothmove}
	d \left( \frac{d_i}{d_j}a_j,\frac{d_i}{d_j}a_k \right)-d \left( \frac{d_{i+1}}{d_j}a_j,\frac{d_{i+1}}{d_j}a_k \right)\leq (d_i-d_{i+1}) d\left( \frac{a_j}{d_j},\frac{a_k}{d_k} \right).
\end{equation}

Define also $f(i)$ as the maximum value of
\begin{align*}
	d(u_1,u_2)+\sum_{j=3}^i\frac{3}{2}d(u_1,u_j)+\sum_{j=3}^i\frac{3}{2}d(u_2,u_j) +\sum_{3\leq j<k\leq i}2d(u_j,u_k)
\end{align*}
over unit vectors $u_1,\ldots,u_i$.

We use equation \eqref{eq: Young} to obtain upper bounds on $f(i)$, which we then use to find numerical values for $B_{n,i}$.

For $i>2$, substituting $\eqref{eq: bothstay}$, $\eqref{eq: onemoves}$, $\eqref{eq: bothmove}$, and  $\eqref{eq: L()}$ into $\eqref{eq: defineDnik}$, we get that
\begin{align*}
	D_{n,i,1} &\leq (d_i-d_{i+1})\frac{f(i)}{n-1}+ (d_i-d_{i+1})\left(\frac{1+3(n-i)+(i-2)+2(n-i)(i-2)}{n-1} \right)\\
	D_{n,i,1} &\leq (d_i-d_{i+1})\left(\frac{f(i)}{n-1}+2\frac{(n-i)(i-1)}{n-1}+1 \right)
\end{align*}
and similarly for $D_{n,i,2}$. So we set
\begin{equation} \label{eq:Bni}
	B_{n,i}=\frac{f(i)}{n-1}+2\frac{(n-i)(i-1)}{n-1}+1
\end{equation}

For $i=2$, the same substition gives us $B_{n,2}=3$. For $i=1$, the same substition gives us $B_{n,1}=1.5$.

If there do not exist four paths $Q_{11}, Q_{12}, Q_{21}, Q_{22}$, then the average length of a path is too great, namely
\begin{align*}
\frac{1}{2} \left( \overline{L(a_{n+1}, \ldots, a_1)}+\overline{L(a_{n+2},a_n,a_{n-1}, \ldots, a_1)} \right) &> r^*S_n\\
\frac{d_{n+1}+d_{n+2}}{2}B_n + \left(d_n-\frac{d_{n+1}+d_{n+2}}{2}\right)B_{n,n}+\sum_{i=1}^{n-1} \left(d_i-d_{i+1}\right)B_{n,i} &> \sum_{i=1}^{n} r^*d_n
\end{align*}

\subsection{$n=3$}
If $d_4 \leq 0.541d_3$, then, by \eqref{eq: Young},
\[\overline{L \left( a_4, a_3, \frac{d_3}{d_2}a_2, \frac{d_3}{d_1}a_1 \right)}\leq 4.677d_3=3r^*d_3.\]
Then, since $B_{3,2}=3<2r^*$ and $B_{3,1}=1.5<r^*$ as in the last section, 
\[\overline{L(a_4, a_3, a_2, a_1)}\leq r^*(d_1+d_2+d_3).\]
Similarly,
\[\overline{L(a_5, a_3, a_2, a_1)}\leq r^*(d_1+d_2+d_3)\]
so the induction works. Thus for $n=3$ we have the constraint
$d_4>0.541d_3$, which is stronger than the one obtained for $n=3$ in the previous section.

\section{Choosing $n$} \label{sec: recombine}
We obtained linear constraints for various values of $n\leq 10$. These, together with the constraints $d_i\geq d_{i+1}$, make a linear program (given in Appendix \ref{sec: lp}), which is unsatisfiable. Thus one of the constraints must not hold, so the induction works for some $n$.

\section{Algorithm}
We repeatedly use the inductive step to obtain a sequence of indices $0=n_0<n_1<n_2<~\ldots$. At stage $j$, we remove $n_j-n_{j-1}$ points.
The intermediate ending points are then of the form $n_j+k_j$ where each $k_j$ is 1 or 2.
Since we are only using $n\leq 10$, we can find all the paths $Q_{11}, Q_{12}, Q_{21}, Q_{22}$ by brute force in linear time.
Now, for both possible values of $k_1$, we find which value of $k_0$ gives the shorter path. Then, for both possible values of $k_2$, we find which value of $k_1$ will make the total path after $a_{n_2+k_2}$ shorter. We repeat until we get to some $n_j>m$, at which point we have two paths and choose the shorter one. This whole algorithm is linear.

\section{Lower bound on degree-3 tree ratios}
\label{sec:lowerbound}
Denote by $\sigma$ the sum of edge weights of the minimal spanning tree and by $\sigma_3$ the sum of edge weights of a minimal degree 3 tree. Denote by $(x_1,x_2,\ldots,x_n)$ the coordinates of a point in $n$ dimensions.

In six dimensions, let $O$ be the origin and let $v_1,v_2,\ldots,v_7$ be the vertices of a simplex with center  at $O$ and radius $\sqrt{6}$.  Let the coordinates of $v_i$ be $(v_{i,1},v_{i,2},v_{i,3},v_{i,4},v_{i,5},v_{i,6})$.  Note that $d(v_i,v_j)=\sqrt{2*7/6}\sqrt{6}=\sqrt{14}$.

Now, given natural $N$ and $0<\alpha < 1$, take the following tree in $7N$ dimensions:
\begin{enumerate}
\item The origin, $O$, is the root.
\item Its $N$ children are $p_1,p_2,\ldots,p_N$.  $p_i$ has coordinates 0 except $x_{7i}=1-\alpha$.
\item Each $p_i$ has seven children, $q_{i,1},q_{i,2},\ldots,q_{i,7}$  The coordinates of $q_{i,j}$ are all 0 except $x_{7i}=1$ and, for $k$ from 1 to 6, $x_{7i-k}=v_{j,k}$.
\end{enumerate}

Then $q_{i,1},q_{i,2},\ldots,q_{i,7}$ form a simplex with center distance $\alpha$ from $p_i$ and with each vertex distance $\sqrt{6}$ from the center.

It is easy to check that
\begin{align*}
d(p_i,p_h)&=\sqrt{2}(1-\alpha) \text{ for } i\neq h\\
d(q_{i,j},q_{i,k})&=\sqrt{14}=d(q_{i,j},q_{h,k}) \text{ for } j\neq k,h\neq i\\
d(p_i,q_{i,j})&=\sqrt{6+\alpha^2}\\
\sigma&=N(1-\alpha+7\sqrt{6+\alpha^2}).
\end{align*}

Then we can pick
\[\alpha=-1-\sqrt{7}+\sqrt{4+4\sqrt{7}},\]
which gives us $d(q_{i,j},q_{h,k})+d(p_i,p_h)=2d(p_i,q_{i,j})$.

Then we can define function $c$ on the vertices so that $c(O)=0, c(p_i)=d(p_i,p_h)/2$ and $c(q_{i,j})=d(q_{i,j},q_{h,k})/2$.
In that case, the length of edge $AB$ is at least $c(A)+c(B)$, so $c$ can be thought of a half-edge length.
Then, since there are $8N+1$ vertices, there are $8N$ edges, so there is a total of $16N$ edge endpoints.  At most 3 of them contribute 0 to $\sigma_3$, at most $3N$ contribute $(1-\alpha)/\sqrt{2}$, and the remainder contribute $\sqrt{14}/2$. Thus 
\[\sigma_3\geq 3N\left(\frac{1}{2}\sqrt{2}(1-\alpha)\right)+(13N-3)\left(\frac{1}{2}\sqrt{14}\right)\]
\[\frac{\sigma_3}{\sigma}=\frac{3N\left(\frac{1}{2}\sqrt{2}(1-\alpha)\right)+(13N-3)\left(\frac{1}{2}\sqrt{14}\right)}{N\left(1-\alpha+7\sqrt{6+\alpha^2}\right)}\]
\[\lim_{N \to \infty}\frac{\sigma_3}{\sigma}=\frac{3\left(\frac{1}{2}\sqrt{2}(1-\alpha)\right)+13\left(\frac{1}{2}\sqrt{14}\right)}{1-\alpha+7\sqrt{6+\alpha^2}}\approx 1.4473\]
Thus $\tau_3^\infty \geq 1.447$.
\section{Acknowledgements}
I thank Samir Khuller for suggesting that I work on this problem and Timothy Chan for improved notation and organization.
\appendix
\section{Values of $x_i$} \label{sec: xis}
\begin{table}[htbp]
\centering
 \label{tbl:fq}
 \caption{$x_j$ values used to bound $f(i)$}
  \begin{tabular}{| c | c | c | }
    \hline
     $i$ & $x_1$ and $x_2$ & $x_3$ through $x_i$ \\ \hline
     3 & 2.127480103088468 & 2.715029663803688 \\ \hline
     4 & 3.2023557495551507 & 4.175556640172782 \\ \hline
     5 & 4.270167577054796 & 5.608618419590356 \\ \hline
     6 & 5.335126162486634 & 7.033301794415261 \\ \hline
     7 & 6.3986555212789265 & 8.454218195486414 \\ \hline
     8 & 7.461367172755974 & 9.873101560726544 \\ \hline
     9 & 8.52356722480373 & 11.290758818589284 \\ \hline
		10 & 9.585425903496056 & 12.707618366991161 \\ \hline
  \end{tabular}
  \end{table}
\begin{table}[htbp]
\centering
 \label{tbl:gq}
 \caption{$x_j$ values used to bound $g(n)$}
  \begin{tabular}{| c | c | c | c | }
    \hline
     $n$ & $x_1$ and $x_2$ & $x_3$ through $x_n$ & $x_{n+1}$\\ \hline
     3 & 2.4556264573869506 & 3.5140460449331314 & 1.5613009117434562 \\ \hline
     4 & 3.5424450202354296 & 4.920230571592636  & 2.294026685501083 \\ \hline
     5 & 4.618609731491003  & 6.336229610465761  & 3.0154193383617174 \\ \hline
     6 & 5.689328832275783  & 7.753335975414664  & 3.7315531287091606 \\ \hline
     7 & 6.757011330006688  & 9.170224016158656  & 4.4448690694127775 \\ \hline
     8 & 7.822844123284092  & 10.58670954888685  & 5.15650608100577 \\ \hline
     9 & 8.88747045789415   & 12.002823667602273 & 5.867063400774457 \\ \hline
		10 & 9.951267362449125  & 13.41863151261787  & 6.576885724382338 \\ \hline
  \end{tabular}
  \end{table} 
  
 \begin{table}[htbp]
 \centering
 \label{tbl:n3}
 \caption{$x_j$ values for the $n=3$ case}
\begin{tabular}{| c | c | c | c | }
    \hline
     $x_1$ & $x_2$ & $x_3$ & $x_4$\\ \hline
     1.2840665853752833 & 1.2840665853752833 & 1.8003074954981302 & 1.0528095728981612 \\ \hline
  \end{tabular}
\end{table}

\section{Linear Program} \label{sec: lp}
This is the infeasible linear program one achieves. The coefficients on the left are strictly greater (by at least 0.0001) than the actual values one would calculate. Since they are all multiplied by poisitive values, this takes care of roundoff error in calculating coefficients.

$\text{Variables: } d_1,\ldots,d_{12}\\
d_i \ge d_{i+1}\text{ for } 1\le i\le 11\\
d_1=1\\
d_4>=0.541d_3\\
4.6568(d_{3}-d_{4})+5.9188(d_{4}-d_{5}/2-d_{6}/2)+3.2034(d_{5}+d_{6})\ge1.559(3d_3+d_{4})\\
4.7426(d_{3}-d_{4})+6.1891(d_{4}-d_{5})+7.3417(d_{5}-d_{6}/2-d_{7}/2)+3.9078(d_{6}+d_{7})\ge1.559(3d_3+d_{4}+d_{5})\\
4.7941(d_{3}-d_{4})+6.3513(d_{4}-d_{5})+7.6734(d_{5}-d_{6})+8.7608(d_{6}-d_{7}/2-d_{8}/2)+4.6125(d_{7}+d_{8})\ge1.559(3d_3+d_{4}+d_{5}+d_{6})\\
4.8285(d_{3}-d_{4})+6.4595(d_{4}-d_{5})+7.8945(d_{5}-d_{6})+9.1341(d_{6}-d_{7})+10.1782(d_{7}-d_{8}/2-d_{9}/2)+5.3177(d_{8}+d_{9})\ge1.559(3d_3+d_{4}+d_{5}+d_{6}+d_{7})\\
4.853(d_{3}-d_{4})+6.5367(d_{4}-d_{5})+8.0525(d_{5}-d_{6})+9.4006(d_{6}-d_{7})+10.5814(d_{7}-d_{8})+11.5946(d_{8}-d_{9}/2-d_{10}/2)+6.0232(d_{9}+d_{10})\ge1.559(3d_3+d_{4}+d_{5}+d_{6}+d_{7}+d_{8})\\
4.8714(d_{3}-d_{4})+6.5946(d_{4}-d_{5})+8.1709(d_{5}-d_{6})+9.6006(d_{6}-d_{7})+10.8837(d_{7}-d_{8})+12.0203(d_{8}-d_{9})+13.0105(d_{9}-d_{10}/2-d_{11}/2)+6.729(d_{10}+d_{11})\ge1.559(3d_3+d_{4}+d_{5}+d_{6}+d_{7}+d_{8}+d_{9})\\
4.8857(d_{3}-d_{4})+6.6397(d_{4}-d_{5})+8.2631(d_{5}-d_{6})+9.7561(d_{6}-d_{7})+11.1189(d_{7}-d_{8})+12.3514(d_{8}-d_{9})+13.4538(d_{9}-d_{10})+14.4259(d_{10}-d_{11}/2-d_{12}/2)+7.4351(d_{11}+d_{12})\ge1.559(3d_3+d_{4}+d_{5}+d_{6}+d_{7}+d_{8}+d_{9}+d_{10})$

\newpage
\singlespacing

\end{document}